\documentclass[aps,prl,twocolumn,showkeys,amsmath,amssymb,longbibliography,floatfix,superscriptaddress]{revtex4-1}
\usepackage[T1]{fontenc}
\usepackage{graphicx}
\usepackage{dcolumn}
\usepackage{amsmath,amssymb,amsfonts,bm,dsfont,units}
\usepackage{hyperref}
\usepackage{tabularx}
\hypersetup{colorlinks=true,citecolor=blue,linkcolor=blue,urlcolor=blue,pdfstartview=FitH}
\usepackage{appendix} 
\usepackage{braket}
\usepackage{float,latexsym}
\usepackage{multirow}
\usepackage{diagbox}
\usepackage[normalem]{ulem} 
\usepackage[caption=false]{subfig} 
\usepackage{comment}
\usepackage{tikz}
\usepackage{xcolor}

\usepackage{array}
\usepackage{graphicx} 
\usepackage{hhline,booktabs}
\usepackage{makecell}
\usepackage{wasysym} 
\usepackage{array}
\newcolumntype{Y}{>{\centering\arraybackslash}X}
\newcommand{\PsiQP}{\Psi_\text{QP}}

\begin{document}

\title{Non-Abelian phases from the condensation of Abelian anyons}

\author{Misha Yutushui}
\affiliation{Department of Condensed Matter Physics, Weizmann Institute of Science, Rehovot 7610001, Israel}
 \author{Maria Hermanns}
 \affiliation{Department of Physics, Stockholm University, AlbaNova University Center, SE-106 91 Stockholm, Sweden}
 \author{David F. Mross}
\affiliation{Department of Condensed Matter Physics, Weizmann Institute of Science, Rehovot 7610001, Israel}
 
\begin{abstract} 
The observed fractional quantum Hall (FQH) plateaus follow a recurring hierarchical structure that allows an understanding of complex states based on simpler ones. Condensing the elementary quasiparticles of an Abelian FQH state results in a new Abelian phase at a different filling factor, and this process can be iterated \textit{ad infinitum}. We show that condensing clusters of the same quasiparticles into an Abelian state can instead realize non-Abelian FQH states. In particular, condensing quasiparticle pairs in the $\nu=\frac{2}{3}$ Laughlin state yields the anti-Pfaffian phase at half-filling. We moreover show that the successive condensation of Laughlin quasiparticles produces quantum Hall states whose fillings coincide with the most prominent plateaus in the first excited Landau level of GaAs.  More generally, such condensation can realize any non-Abelian FQH state that admits a parton representation. This surprising result is supported by an exact analysis of explicit wavefunctions, field theory arguments, conformal-field theory constructions of trial states, and numerical simulations. 
\end{abstract}

\date{\today}
\maketitle

 The FQH effect has significantly shaped the understanding of quantum many-body states~\cite{Tsui_fqh_1982}. Its study introduced fundamental concepts such as chiral edge states~\cite{Wen_CLL_edge_1990}, topological order~\cite{Wen_topological_1990}, and fractionalization~\cite{Laughlin_fqh_1983,Haldane_fqh_1983,Halperin_QH_1983} in an experimentally accessible platform. In particular, Abelian FQH states at the filling factor $\nu=\frac{q}{p}$ support at least $p$ quasiparticle types, which differ in charge or statistics. Among them, two quasiparticles assume a special role. (i) Fundamental quasiparticles carry the minimal non-zero fractional charge $e_* = \frac{1}{p}e$ permitted by the FQH state. (ii) Laughlin quasiparticles carry the charge $e_\Phi = \nu e$ associated with inserting one magnetic flux quantum. Both types of quasiparticles are routinely observed in transport experiments. At $\nu=\frac{1}{3}$, fundamental and Laughlin quasiparticles coincide; they have been observed via shot noise~\cite{Saminadayar_fractional_charge_1997,Picciotto_fractional_charge_1998} and, more recently, in interference experiments~\cite{Nakamura_Aharonov_2019,Nakamura_Direct_2020,Kundu_anyonic_2023,Kim_Aharonov_2024,werkmeister2024anyon,samuelson_anyonic_2024}. Similar measurements at other fillings \cite{Chung_Scattering_2003,Dolev_observation_2008,Bid_Shot_Noise_2009,Dolev_Dependence_2010,Biswas_Shot_2022,Nakamura_Fabry_2023,Ghosh_OMZI_2024,Ghosh_Coherent_Bunching_2024,Kim_Aharonov_even_2024} have observed both types of fractional excitations, with Laughlin quasiparticles dominating at the lowest temperatures.

\begin{figure}[tbh]
 \centering
 \includegraphics[width=0.99
 \linewidth]{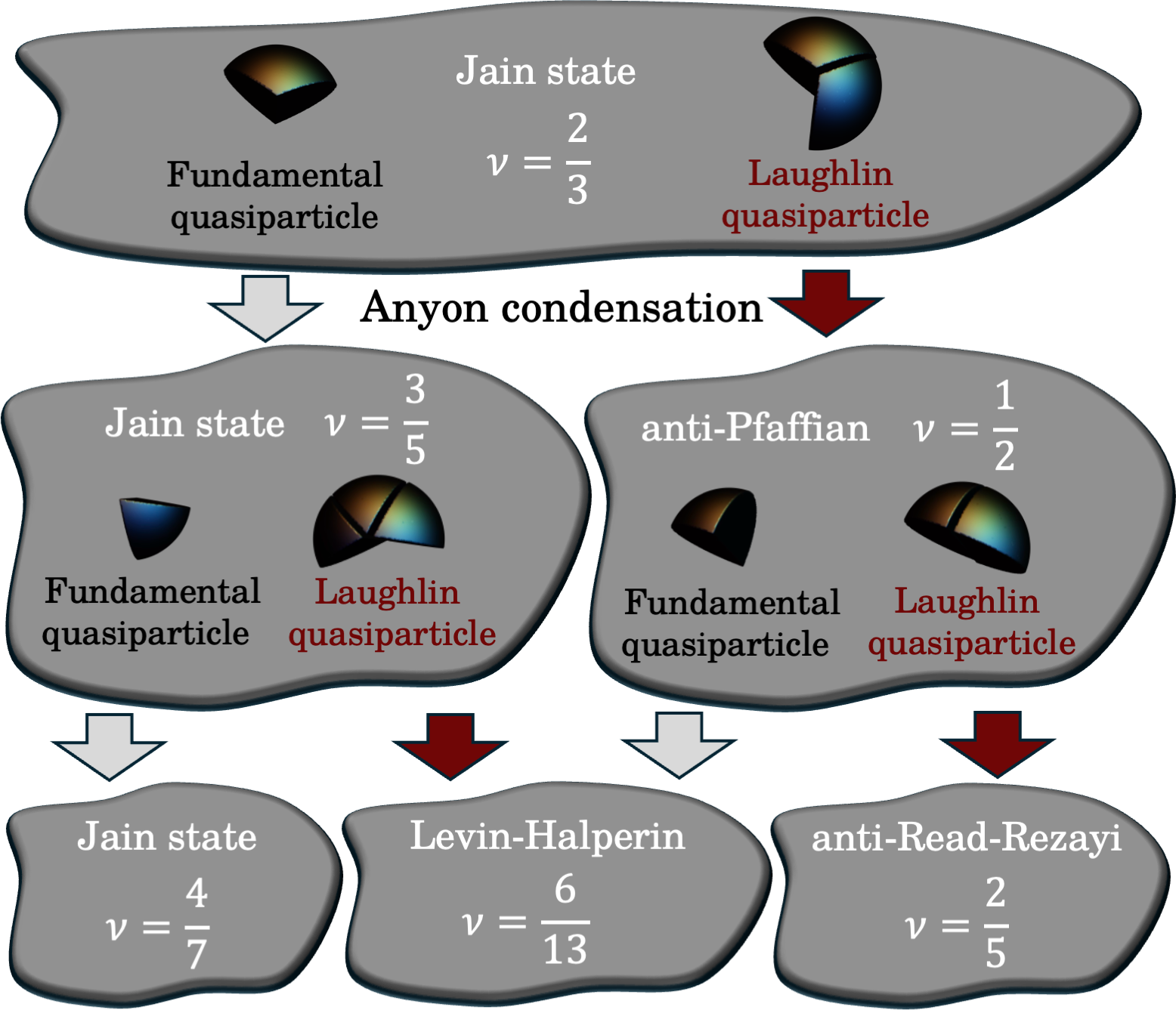}\\
 \caption{{\bf Anyon condensation pattern} The fundamental quasiholes of the $\nu=\frac{2}{3}$ Jain state carry charge $e/3$, while Laughlin quasiholes carry $2e/3$. Condensing them yields the Abelian $\nu=\frac{3}{5}$ state and the non-Abelian anti-Pfaffian with $\nu=\frac{1}{2}$, respectively. Further quasiparticle condensation yields the $\nu=\frac{4}{7}$ Jain state, the $\nu=\frac{6}{13}$ Levin-Halperin state, or the non-Abelian anti-Read-Rezayi state at $\nu=\frac{2}{5}$.}
 \label{fig.anyone}
\end{figure}
Fractional quasiparticles of either type can be created when the filling factor of the 2DEG deviates from the rational number corresponding to the Hall conductance $\sigma_{xy}$. The observed plateaus in $\sigma_{xy}$ occur if the excess quasiparticles become localized and do not participate in electrical transport. Alternatively, fractional quasiparticles can themselves form a quantum Hall state, resulting in a new plateau with a different Hall conductance. This \textit{hierarchical} perspective on FQH states well explains the systematic pattern observed plateaus in the lowest Landau level (LLL). Its formulations using explicit trial wavefunctions~\cite{Haldane_fqh_1983,Halperin_Statistics_1984} or topological field theory \cite{Wen_Topological_1995} are known as \textit{anyon condensation}. Different implementations of Abelian anyon condensation have consistently found that the resulting FQH states are `no greater than the sum of their parts.' In particular, condensing fundamental Abelian quasiparticles in an Abelian FQH state results in new Abelian topological orders. 

Non-Abelian FQH states exhibit even more exotic properties~\cite{Moore_nonabelions_1991,Greiter_half_filled_1991,Wen_Non-Abelian_1991,Read_Beyond_1999}. In particular, non-Abelian quasiparticles imply a manifold of topologically degenerate ground states~\cite{Oshikawa_Topological_2007}. A particular state in this manifold can serve as a robust quantum memory upon which braiding processes act as quantum gates; see ~\cite{Stern_non_Abelian_2010,nayak_non-abelian_2008} for two relevant reviews. The best-known example of such a phase is the Moore-Read Pfaffian \cite{Moore_nonabelions_1991}. It is one of the primary candidates for explaining the $\nu=\frac{5}{2}$ plateau in GaAs and several half-filled plateaus in graphene~\cite{Kumar_Quarter_2024,Ki_bilyaer_graphene_2014,Li_bilayer_graphene_2017,Zibrov_Even_Denominator_2018,Zibrov_Tunable_bilayer_graphene_2017,Kim_Even_Denominator_f_wave_2019,Huang_Valley_bilayer_graphene_2022,chen_tunable_2023}. Different non-Abelian states also follow hierarchical patterns~\cite{Bonderson_hierarchy_2008,Levin_collective_2009,Hermanns_condensing_2010,yutushuidaughter2024,Zheltonozhskii_daughters_2024,zhang_hierarchy_2024}, which include states whose Fibonacci anyons are capable of universal quantum information processing~\cite{nayak_non-abelian_2008,Stern_non_Abelian_2010}.

In this work, we demonstrate that numerous non-Abelian orders can arise from the condensation of Laughlin quasiparticles (CLQ) in Abelian FQH states. An important example of this effect is illustrated in Fig.~\ref{fig.anyone} for the $\nu=\frac{2}{3}$ state. 
While fundamental quasiparticle condensation leads to an Abelian Jain state \cite{Jain_composite-fermion_1989,Jain_composite_2007}, the condensation of Laughlin quasiparticles---a composite of two fundamental ones---yields a non-Abelian phase known as anti-Pfaffian \cite{Levin_particle_hole_2007,Lee_particle_hole_2007}. CLQ at the anti-Pfaffian parent state, surprisingly, yields anti-Read-Rezayi topological order.

Remarkably, the most prominent FQH states in the second Landau level (SLL) of GaAs occur at the same partial fillings $\nu=\frac{1}{3},\frac{2}{5},\frac{1}{2},\frac{2}{3}$ as arise from successive CLQ at $\nu=1$ (quasihole condensation) or at $\nu=\frac{1}{3}$ (quasielectron condensation). In the main text, we focus primarily on quasiholes and discuss quasielectron condensation via conformal field theory analysis in the End Matter; see also Appendix~\ref{app.SU22} for numerical simulations.  

\textit{Consecutive CLQ at $\nu=1$}---The wavefunction of a single filled Landau level with $N_e$ electrons on a disk is given by $\chi_1 = \prod^{N_e}_{i<j}(z_i -z_j)$, where $z_i = x_i + i y_i$ are complex coordinates, and we omitted a Gaussian factor. At  $\nu=1$, fundamental and Laughlin quasiholes coincide and carry charge $-e$. They can be created by reducing $N_\text{e}$ or increasing the magnetic flux $N_\Phi$. In our analysis, we fix $N_\text{e}$ and introduce quasiholes or quasielectrons by increasing or decreasing $N_\Phi$, respectively. Creating a hole at the complex coordinate $u_A$ amounts to multiplying the \textit{parent state} $\chi_1$ by the factor $\prod_i(z_i -u_A)$. Introducing $N_\text{QP}$ of such excitations and placing them into a Laughlin state yields a new \textit{descendant} wavefunction
\begin{align}
 \Psi_\text{CLQ} (\{z_i\}) =\chi_1 \underbrace{\prod\limits_{i=1}^{N_e}\prod\limits_{A<B}^{N_\text{QP}}\int_{u_A} (u_A-z_i) (u^*_A-u^*_B)^{2}}_{\equiv P(\{z_i\})}.\label{eqn.iqhanyoncondense}
\end{align}
The polynomial factor $P(\{z_i\})$ is holomorphic and symmetric in all its arguments. Moreover, it satisfies $P(\{z_i\})=P(\{z_i + a\})$ and thus describes a homogeneous quantum Hall state on an infinite disk. On a sphere, one replaces all differences $z_i -z_j$ with appropriate spinor coordinates \cite{Haldane_fqh_1983}.

The integral in Eq.~\eqref{eqn.iqhanyoncondense} is non-zero only when the powers of $u$ in the first term compensate for those of $u^*$ in the second term.  This requirement fixes the number of quasiholes as $N_\text{QP} = \frac{N_\text{e}}{2}+1$ for disk and sphere geometry, which implies even $N_e$. Each Laughlin quasihole is associated with one flux quantum. Together with the $N_e-1$ flux quanta of $\chi_1$, we obtain the total magnetic flux of $\Psi_\text{CLQ}= \chi_1 P$ as $N_{\Phi,\text{CLQ}}= \frac{3}{2}N_\text{e}$, identifying its filling factor to be $\nu=\frac{2}{3}$. 

In the discussion so far, the parent state acted as a mere spectator, and $\chi_1$ in Eq.~\eqref{eqn.iqhanyoncondense} could be replaced by any LLL trial state. In particular, it could be replaced by $\Psi_\text{CLQ}$ obtained from a previous CLQ generation. Iterating the Laughlin-quasihole condensation in this way yields a sequence of wavefunctions, $\Psi_\text{CLQ}^{(k)} =\Psi_\text{CLQ}^{(k-1)} P=\Psi_\text{CLQ}^{(k-2)} P P=\ldots=\chi_1 P^k$, 
at the magnetic flux
\begin{align}
 N_{\Phi,\text{CLQ}} = \frac{2+k}{2} N_e -(1-k).
\end{align}
The coefficient before $N_e$ identifies the filling factor as $\nu_\text{CLQ}=\frac{2}{2+k}$ and the constant term the shift as ${\cal S}_\text{CLQ} = 1-k$~\cite{Wen_shift_1992}. These are the quantum numbers of the non-Abelian anti-Read-Rezayi (aRR) sequence \cite{Read_Beyond_1999}, which requires even $N_e$, consistent with our construction. Surprisingly, the wavefunctions $\Psi^{(k)}_\text{CLQ}$ for arbitrary $k$ were obtained using only Abelian ingredients at each step. Do they still describe the non-Abelian phases as indicated by $\nu$ and ${\cal S}$?

To confirm the topological order of $\Psi^{(k)}_\text{CLQ}$, we explicitly express it as the particle-hole conjugate of a Read-Rezayi state. We begin by examining the $k$th power of the polynomial $P$, i.e.,
\begin{equation}
 \begin{split}
 P^k &=
\prod\limits_{i=1}^{N_e}\prod_{\sigma=1}^k \prod\limits_{A}^{N_\text{QP}}\int_{u_{A,\sigma}}(u_{A,\sigma}-z_i)\prod\limits_{B<A}^{N_\text{QP}}(u^*_{A,\sigma}-u^*_{B,\sigma})^2. 
 \end{split}
 \label{eqn.pk}
\end{equation}
Under the products over $\sigma$ and $A$, the first term of the integrand is invariant under any permutation of the quasihole coordinates $u_{A,\sigma}$, including those with different $\sigma$. Consequently, only the symmetric part of the second term gives a non-zero contribution to the integral. The symmetrization
\begin{align}
 {\cal S}\prod_{\sigma=1}^{k}\prod_{A<B} (u^*_{A,\sigma}-u^*_{B,\sigma})^2 = \Psi^{(k)}_\text{bRR}(\{u^*\})
 \label{eqn.symm}
\end{align}
yields the bosonic Read-Rezayi wavefunction at $\nu=\frac{k}{2}$ \cite{Regnault_Bridge_2008,Barkeshli_projective_2010,Cappelli_Parafermion_2001}, which is non-Abelian for $k>1$. Using Eqs.~\eqref{eqn.pk} and \eqref{eqn.symm}, we now express $ \Psi_\text{CLQ}^{(k)}$ as the explicit particle-hole conjugate~\cite{Girvin_Particle_hole_1984} of a fermionic Read-Reazyi wavefunction, i.e.,
\begin{align}
 \Psi_\text{CLQ}^{(k)}(\{z\})=\int_{\{u\}}\chi_1(\{z,u\})\left[\tilde\Psi^{(k)}_\text{RR}(\{u\})\right]^*~.
 \label{eqn.phconj}
\end{align}
Here, $\chi_1(\{z,u\})$ describes a $\nu=1$ state for all $N_e + k N_\text{QP}$ coordinates $z_i$ and $u_{A,\sigma}$. The fermionic Read-Rezayi wavefunction in Eq.~\eqref{eqn.phconj} only differs by a topologically trivial factor $|\chi_1|^2$ from the original Read-Rezayi wavefunction; see Appendix~\ref{app.psitildefermion}. Consequently, sequential CLQ at $\nu=1$ generates the non-Abelian aRR sequence.

This analytical result relies on a specific form of $\Psi_\text{CLQ}^{(k)}$, which exhibits an enhanced symmetry due to $k$ identical factors of $P$. However, topological orders should be insensitive to microscopic details. In particular, we expect CLQ in any parent wavefunction realizing a given phase to yield the same descendant topological order. To test this expectation numerically, we analyzed CLQ using Eq.~\eqref{eqn.iqhanyoncondense} with $\chi_1$ replaced by Jain's composite-fermion wavefunction for $\nu=\frac{2}{3}$ ($k=1$) \cite{Jain_composite_2007}. Similarly, for CLQ in the anti-Pfaffian at $\nu=\frac{1}{2}$ ($k=2$), we replaced $\chi_1$ by $\psi_\text{aPf} = {\cal P}_\text{LLL} (z_i -z_j)^2 \text{Pf}\left[\frac{z_i-z_j}{(z_i^*-z_j^*)^2}\right]$ \cite{Yutushui_Large_scale_2020}. 

\begin{figure} [h] \centering
 \includegraphics[width=0.99
 \linewidth]{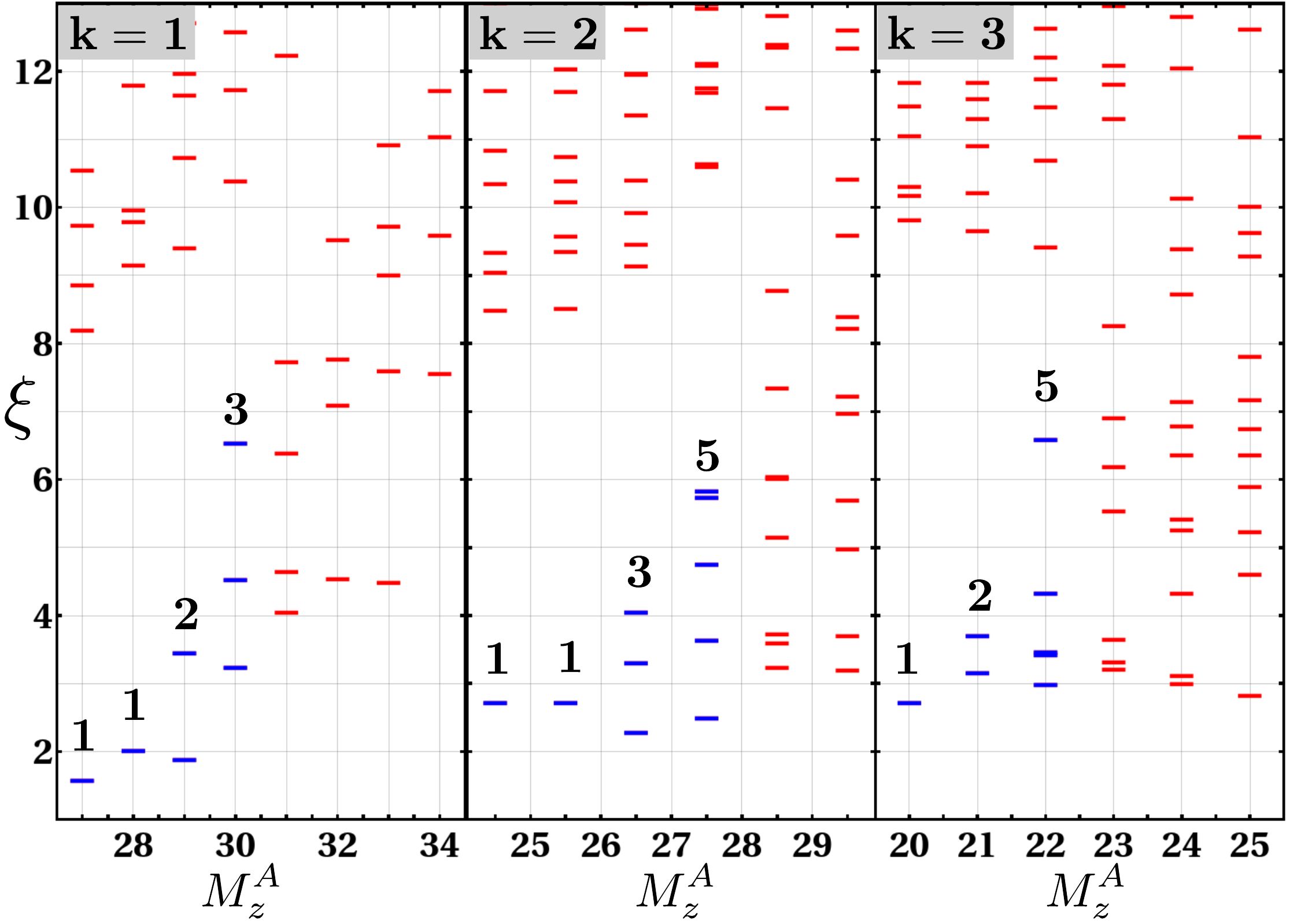}\\
 \caption{{\bf Orbital entanglement spectra} We show the OES of $\Psi_\text{CLQ}^{(k)}$ with $k=1,2,3$ for the largest electron numbers $N_e$ in Table~\ref{tab.CLQ}. Specifically, we compute the eigenvalues of the reduced density matrix for $N_e/2$ electrons, which all have positive angular momenta. At these particle numbers, the OES are expected to match trivial sectors for Jain and anti-Pfaffian states and the parafermion sector of anti-Read-Rezayi. The pattern of low-lying states (indicated in blue) agrees with the expectation for these topological orders.}
 \label{fig.ES}
\end{figure}

We obtained a Fock-space representation of the resulting $\Psi_\text{CLQ}^{(k)}$ for $k=1,2,3$ via a brute-force Monte-Carlo approach, computing the overlaps with all many-body basis states of $N_\text{e}=6-12$ electrons in the LLL on a sphere (see Appendix~\ref{app.MC_details} for details). We then calculated overlaps with level-$k$ aRR states obtained using Jack-polynomials \cite{Bernevig_Anatomy_2009}. In all cases, we find squared overlaps above 98$\%$; see Table~\ref{tab.CLQ}. Fig.~\ref{fig.ES} shows the corresponding orbital entanglement spectra (OES) \cite{Li_entanglement_spectrum_2008}, whose low-lying states follow the known counting of the aRR level-$k$ topological order.

\textit{CLQ in general FQH states}---We now generalize Eq.~\eqref{eqn.iqhanyoncondense} to condense Laughlin quasiholes of any single-component parent state $\Psi$ into a chosen FQH 
state. The resulting wavefunction is
\begin{align}
 \Psi_\text{CLQ} (\{z_i\}) =\Psi(\{z_i\}) \prod_{A=1}^{N_\text{QP}}\int_{u_A} (u_A-z_i)\PsiQP^* (\{u_A\}),\label{eqn.anyoncondense}
\end{align}
where the \textit{pseudo-wavefunction} $\PsiQP$ describes the state formed by the anyons. When $\PsiQP$ is a bosonic hierarchy, and $\Psi$ is a Laughlin state, one recovers the Haldane hierarchy of Ref.~\onlinecite{Haldane_fqh_1983}. To condense quasielectrons instead of quasiholes, one must take the complex conjugate of the integral in Eq.~\eqref{eqn.anyoncondense} and project the resulting wavefunction to the LLL. A further generalization, where clusters of $q$ Laughlin quasiparticles are condensed by raising the factor $(u_\alpha-z_i)$ to the $q$th power, is discussed in Appendix~\ref{app.cluster}.

The number of charge-$e_\Phi$ quasiparticles required to form a homogenous FQH state is $
N_\text{QP}~=~\nu_\text{QP} ( N_\text{e}+{\cal S}_\text{QP})$,
where $\nu_\text{QP}$ and ${\cal S}_\text{QP}$ are the filling factor and shift of $\PsiQP$. Each power of $z$ corresponds to one flux quantum, which implies
\begin{align}
 \nu_\text{CLQ} = \frac{\nu}{1 \pm \nu \nu_\text{QP}}~,
 \qquad
 {\cal S}_\text{CLQ}={\cal S} \mp {\cal S}_\text{QP} \nu_\text{QP}~,
 \label{eqn.fillingshift}
\end{align}
in a straightforward generalization of Ref.~\onlinecite{Haldane_fqh_1983}. The top sign holds for quasihole condensation, and the bottom sign is for quasielectrons. In contrast to fundamental quasiparticle condensation, the condensation of Laughlin quasiholes or quasiparticles produces a single state at each hierarchy level.The wavefunction in Eq.~\eqref{eqn.anyoncondense} can exhibit a non-integer shift, e.g., when quasiparticles are condensed into a bosonic Jain state at $\nu=\frac{n}{np+1}$ with $|n|,|p|>1$, allowing access to states beyond the reach of standard composite fermion and parton constructions.

\begin{table}[h]
	\centering
	\renewcommand{\arraystretch}{1.2} 
	\caption{{\bf Overlaps of CLQ wavefunctions} Squared overlaps of trial states obtained via CLQ with model wavefunctions of the $\nu=\frac{2}{3}$ Jain state, $\nu=\frac{1}{2}$ anti-Pfaffian, and the $\nu=\frac{2}{5}$ anti-Read-Rezayi state. For perspective, the squared overlap between anti-Pfaffian and composite Fermi liquid with $N_e=10$ is 0.38(4$\pm$2). }
	\begin{tabularx}{\linewidth}{c | Y Y Y}
 \hline\hline 
 &\multicolumn{3}{c}{$|\langle \text{Anyon condensation} | \text{Model wavefunction} \rangle|^2$} \\
 		 &  Jain state& 
                 anti-Pfaffian & 
                 anti-Read-Rezayi 
		\\
		 $N_{e}$ & 
 $k=1$ $(\nu=\frac{2}{3})$& 
 $k=2$ $(\nu=\frac{1}{2})$ & 
 $k=3$ $(\nu=\frac{2}{5})$ 
		\\ \hline
		6 & 0.99643(0$\pm$1) & 0.99034$(8\pm1)$ & 0.9977(2$\pm$1) \\
		8 & 0.9954(4$\pm$1) & 0.990$(6\pm1)$ & 0.99(6$\pm$5) \\
		10 & 0.993(6$\pm$3) & 0.98$(1\pm5)$ & --- \\
		12 & 0.99$(2\pm6)$ & --- & --- \\
		\hline\hline
	\end{tabularx}
	\label{tab.CLQ}
\end{table}

According to these quantum numbers, successive Laughlin quasielectron condensation at $\nu=\frac{1}{3}$ yields a sequence of topological orders that mirrors Fig.~\ref{fig.anyone}. The first two members of this sequence are the Abelian $\nu=\frac{2}{5}$ Jain state and the non-Abelian SU(2)$_2$ state at $\nu=\frac{1}{2}$ \cite{Wen_Non-Abelian_1991}. We confirm this identification by numerically computing the overlap between states obtained via any condensation and known SU(2)$_2$ trial states, finding squared overlaps above $0.98$ for $N_e\leq 12$. Given the large overlaps between these states, we expect their microscopic properties, such as quasiparticle energies or the edge modes velocities in the presence of a confining potential, also to be similar.

\textit{Parton interpretation}---To provide additional insights into the origin of the non-Abelian topological orders, we relate them to the `parton' approach to FQH states \cite{Jain_Incompressible_1989}. This framework obtains FQH wavefunctions as the product of integer quantum Hall wavefunctions $\chi_n$ describing $n$ filled Landau levels. In particular, the $\nu=\frac{2}{3}$ Jain state is described by $\Psi^\text{parton}_{2/3} = {\cal P}_\text{LLL}\chi_1^2 \chi_{2}^*$. Comparing it to Eq.~\eqref{eqn.iqhanyoncondense} with $k=1$, we conclude that 
\begin{align}
P(\{z_i\})\sim \chi_1 \chi_{2}^*\label{eqn.1parton} .
\end{align}
The two sides of this equation are, of course, not identical. In particular, the left-hand side is holomorphic, while the right-hand side is not. Still, 
 they are expected to realize identical topological orders once multiplied with a suitable wavefunction and projected to the LLL. The polynomial $P(\{z_i\})$ is already manifestly in the LLL and represents a multiplicative factor that is independent of the parent state. Using \eqref{eqn.1parton}, we obtain
\begin{align}
\Psi_{\frac{2}{2+k}}\equiv \chi_1 P(\{z_j\})^k \sim {\cal P}_\text{LLL}\chi_1^{1+k}(\chi_{2}^*)^k~,\label{eqn.aRRseries}
\end{align}
 a known parton representation of the aRR states containing anti-holomorphic SU(2)$_{k}$ topological order \cite{Balram_parton_2018,Balram_parton_2019}. 

We now consider Eq.~\eqref{eqn.anyoncondense} for cases where the pseudo-wavefunction $\PsiQP^*$ describes a bosonic Jain state at $\nu_\text{QP} =\frac{n}{n+1}$. Following the same logic as in the previous example, we conjecture that 
\begin{align}\label{eq.boson-vortex}
\prod_{\alpha=1}^{N_\text{QP}}\int_{u_A} (u_A-z_i)\chi_1(\{u_A^*\})\chi_n(\{u_A^*\}) \sim \chi^*_{n+1} \chi_1
\end{align}
holds irrespective of a specific trial state it multiplies. The wave function on the right hand side of Eq.~\eqref{eq.boson-vortex} has filling factor $\nu_\text{QP}^{-1}$, which is a manifestation of boson-vortex duality~\cite{Fisher_Correspondence_1989}. For the cases where the resulting wavefunctions are Jain states, this equivalence was established explicitly in Ref.~\cite{Read_Excitation_1990} and confirmed numerically for two examples in Ref.~\cite{yang_hierarchy_1994}.
The validity of this identity---which we have not proven in general---would imply that \textit{any} parton state can be obtained by successive condensation of Abelian anyons based on Laughlin states.

\textit{Relation to field theoretical anyon condensation}---
It is illuminating to view our results in the context of two field-theoretical techniques. The K-matrix approach developed in Ref.~\cite{Wen_Topological_1995} is restricted to Abelian parent topological orders and results in Abelian descendants. To reconcile it with our findings, we note a crucial difference between the two approaches: The explicit anyon condensation yields single-component wavefunctions antisymmetric in all coordinates. In contrast, K-matrices generically translate into multi-component states. It is well known that (anti-)symmetrizing a wavefunction can change its topological order, e.g., Eq.~\eqref{eqn.symm}. Our analytical expression in Eq.~\eqref{eqn.phconj} made explicit use of this relation, which suggests single-component-ness as the origin of non-Abelian statistics.

Ref.~\onlinecite{zhang_hierarchy_2024} developed a novel perspective on quantum Hall hierarchies that captures arbitrary topological orders. There, hierarchical states are obtained by stacking a second FQH state with a different filling factor onto the parent. A charge-neutral bosonic quasiparticle constructed by combining anyons of the two layers is then identified and condensed.

Obtaining a half-filled state from a parent $\nu=\frac{1}{3}$ Laughlin state requires stacking with $\nu_\text{Stack}=\frac{1}{6}$. In the spirit of the present article, we choose an Abelian state, specifically the strong-pairing state described by $K=24$. Combining an elementary quasiparticle of the Laughlin state with an $e/3$ excitation of the strong pairing state yields a \textit{neutral fermion} $\psi$. Being a fermion, $\psi$ can only condense in pairs. The statement $\langle\psi^2 \rangle \neq 0$ does not fully characterize the resulting phase, which additionally requires the specification of its Symmetry Protected Topological (SPT) nature. Depending on this choice, the resulting half-filled state is either Abelian or non-Abelian. 

We conjecture that the wave-function-based approach naturally selects a non-trivial SPT with an odd Chern number due to its single-component nature. By contrast, the K-matrix approach corresponds to an even Chern number. It would be interesting to understand how this choice of SPT can be changed in either approach, but such an analysis is beyond the scope of the present study.

\textit{Discussion}---We have demonstrated at the level of explicit wavefunctions that the condensation of Abelian Laughlin quasiparticles can generate non-Abelian topological orders. In particular, we analytically established the emergence of the anti-Read-Rezayi state by sequentially condensing Laughlin quasiholes of a parent $\nu=1$ state. Our numerical results show that the topological phase of the descendent is insensitive to the microscopic details of the parent or the specific implementation of CLQ. Based on these findings, we conjecture that \textit{the topological phase described by the product of two wavefunctions is determined solely by its contituents' topological orders}.

In particular, we observe that, according to Eq.~\eqref{eqn.fillingshift}, CLQ at $\nu=\frac{3}{5}$ leads to a state with $\nu=\frac{6}{13},{\cal S} =-2$---the same quantum numbers as a Levin-Halperin daughter of the anti-Pfaffian \cite{Levin_collective_2009}. Similarly, CLQ at $\nu=\frac{3}{7}$ yields a daughter of the SU(2)$_2$ \cite{yutushuidaughter2024,Zheltonozhskii_daughters_2024}. These phases were originally accessed by condensation of non-Abelian anyons. Obtaining them via Abelian CLQ suggests an underlying structure that could potentially be used as a shortcut to descendants of other non-Abelian states. 

Our findings have interesting ramifications for two famous FQH states in the SLL of GaAs. The $\nu=2+\frac{1}{2}$ plateau is thought to be non-Abelian, with numerical studies indicating Moore-Read or anti-Pfaffian orders. CLQ in these two states yields the two most prominent non-Abelian candidates at $\nu=2+\frac{2}{5}$. Firstly, the $k=3$ aRR state, and secondly, the Bonderson-Slingerland state \cite{Bonderson_hierarchy_2008}. This concurrence suggests a more detailed look into these phases, in particular, the energetics favoring either one. Additionally, the experimental evidence of a different $\nu=2+\frac{1}{2}$ topological order \cite{Banerjee_observation_2018,Dutta_novel_2022} suggests even richer possibilities and a potentially interesting role of the disorder \cite{Mross_theory_2018}.

More generally, the observed plateaus in the LLL and SLL of GaAS are consistent with the condensation of fundamental and Laughlin quasiparticles, respectively. Could the different energetics of either quasiparticle type in the two Landau levels explain the realized FQH states? In particular, this hypothesis would predict that Laughlin quasiparticles are more prominent in the SLL and fundamental quasiparticles in the LLL. This property could be tested numerically or in experiments sensitive to the charge of individual quasiparticles.

\begin{acknowledgments}
\textit{Acknowledgments}---
It is a pleasure to acknowledge illuminating discussions with Biao Lian, Xiao-Gang Wen, Eddy Ardonne, Thors Hans Hansson, and Steven Simon. This work was supported by the Israel Science Foundation (ISF) under grant 2572/21 and by the Minerva Foundation with funding from the Federal German Ministry for Education and Research. MH acknowledges funding by the Knut and Alice Wallenberg Foundation under grant no. 2017.0157.
\end{acknowledgments}

 \section*{End Matter}
 \label{app.cft}

 \subsection*{From Laughlin $\nu=\frac{1}{3}$ to $\nu=\frac{1}{2}$ SU(2)$_2$---an explicit construction}
The main text focuses on non-Abelian topological orders arising from the consecutive condensation of Laughlin quasiholes. Here, we provide an explicit example of analogous physics in the case of quasielectron condensation. We employ conformal field theory techniques to demonstrate that the successive condensation of Laughlin quasielectrons at $\nu=\frac{1}{3}$ yields first the $\nu=\frac{2}{5}$ Jain and second the $\nu=\frac{1}{2}$ SU(2)$_2$ order. 
The first step reproduces the known results of Ref.~\cite{Hansson2009quantum} using the framework of Ref.~\cite{hansson2017quantum}. Further technical details can be found in Refs.~\cite{Hansson2009quantum, Henderson_conformal_2024} whose conventions we adopt.

\paragraph{Quasielectron condensation at $\nu=\frac{1}{3}$}
The Laughlin state at $\nu=\frac{1}{3}$, $\Psi_{\text {Laughlin}}= \prod_{i<j}(z_i-z_j)^3$,
can be expressed as the correlation function of holomorphic vertex operators $V(z)$, i.e.,
 \begin{align}
 \Psi_{\text {Laughlin}}&= \langle \prod_{j=1}^N V(z_j)\rangle \,.
 \label{eqn.Laughlin.vertex}
 \end{align}
The conventional choice is $V(z)=e^{i \sqrt{3} \phi(z)}$, with a chiral boson $\phi(z)$ satisfying 
$\langle e^{i \alpha \phi(z)}e^{i \alpha \phi(w)} \rangle \propto (z-w)^{\alpha^2}$. 
However, for anyon condensation, it is more convenient to instead factorize the vertex operator into a flux attaching factor $e^{i \sqrt{2}\phi(z)}$ and a composite fermion factor $e^{i \phi_1(z)}$, where $\phi_1$ is a second chiral boson with the same correlation function as $\phi$. We denote this modified vertex operator by
\begin{align}\label{eq:V1}
 V^{(1,+)}(z)&\equiv e^{i \sqrt{2}\phi(z)}e^{i \phi_1(z)}, 
\end{align}
which can be equivalently used in Eq.~\eqref{eqn.Laughlin.vertex}.

 The vertex operator of the Laughlin quasihole can be written as 
 \begin{align}\label{eq:qh}
 H_L(u)=e^{i\phi_1(u)}e^{-i \phi_2(u)}. 
 \end{align}
When quasiholes are inserted into the correlator $\langle V^{(1,+)}(z_1) V^{(1,+)}(z_2)\ldots \rangle$, the first exponential yields the $(z_i-u_A)$ factor of Eq.~\eqref{eqn.iqhanyoncondense} in the main text. 
The second exponential ensures that the $(u_A-u_B)$ factor appears with the second power, allowing us to use bosonic pseudo-wavefunctions, as in the main text. 

Quasielectrons are more subtle. Naively representing them as inverse quasiholes, $H^{-1}(u)$, leads to singularities in the wavefunction.
Ref.~\cite{Hansson2009conformal,Hansson2009quantum} resolved this problem by defining the action of a regularized inverse hole as

\begin{align}\label{eqn.reg.inverse.hole}
 &\langle H^{-1}(u) \underbrace{V \ldots V}_{N_e}\rangle\equiv 
 \sum_{\alpha=1}^{N_e} (-1)^\alpha f(z_\alpha-u)\langle V'(z_\alpha) \underbrace{V \ldots V}_{N_e-1}\rangle
\end{align}
where $f(z_\alpha-u)$ is 
a Gaussian factor (up to a phase) and
\begin{align}
 V'(z)&= \partial_z :H^{-1}V:(z). \label{eqn.inv.hole.dresses.electron}
\end{align}
The derivative is necessary to obtain a non-vanishing result and reflects the need to promote one composite fermion into a higher $\Lambda$ level when reducing the flux, thus leading to an additional factor $\bar z\sim \partial_z$. 
For the Laughlin state, the altered vertex operator is given by 
 \begin{align}\label{eq:V2}
 V^{(2,+)}(z)&= \partial_z : H^{-1} V^{(1,+)}:(z)= \partial_z e^{i\sqrt{2}\phi(z)} e^{i\phi_2(z)}. 
 \end{align}

A surprising benefit is that the integration over the quasielectron coordinates can now be performed analytically---the Gaussian factors combined with the exponential factors from the Landau levels become the lowest Landau level delta functions, and the integration becomes a simple summation over all possible permutations of attaching the $N_\text{QP}$ holes to the $N_e$ electrons. 
For $N_\text{QP}=N_e/2$ \footnote{On the disc geometry, the additional '-1' used in the main text is a boundary effect and, thus, unimportant for the topological features. For a proper construction on the sphere, the reader is referred to Ref.~\cite{Kvorning_quantum_2013}.} quasiparticles, one recovers the well-known $\nu=\frac{2}{5}$ hierarchy/Jain state: 
 \begin{align}
 \label{eq:2/5}
 \Psi_{2/5}= \mathcal{A} \langle \underbrace{V^{(1,+)}\ldots V^{(1,+)}}_{N_e/2} \underbrace{ V^{(2,+)} \ldots V^{(2,+)}}_{N_e/2}\rangle
 \end{align}
 where $\mathcal{A}$ denotes antisymmetrization. The antisymmetrization can be incorporated by imposing anti-commutation relations of the vertex operators \cite{Henderson_conformal_2024}.

By combining both vertex operators into a single electron operator, 
\begin{align}
 V_e(z) &= (e^{i \phi_1(z) }+ \partial e^{i \phi_2(z)}) e^{i \sqrt{2} \phi(z)}.
 \label{eqn.2over5.electron},
 \end{align}
  the $\nu=\frac{2}{5}$ wavefunction can be succinctly expressed as a single correlator,
in complete analogy to the Laughlin state
 \begin{align}
	\Psi_{2/5}&= \langle \prod_{j=1}^N V_e(z_j)\rangle . \label{eqn.cft.2over5} 
\end{align}

\paragraph{Laughlin quasielectron condensation at $\nu=\frac{2}{5}$}
 The $\nu=\frac{2}{5}$ state supports two fundamental quasiholes, $H_1(u)=e^{i \phi_1(u)}$ and $H_2(u)=e^{i \phi_2(u)}$, whose combination is the Laughlin quasihole: 
 \begin{align}
 H_L(u)=H_1(u)H_2(u).
 \end{align}
Condensing the quasielectron connected to $H_2$ \footnote{As in Eq.~\eqref{eq:qh}, one first introduces a new chiral boson $\phi_3$ to condense into a bosonic pseudo-wavefunction.} yields the standard hierarchy state at filling $\nu=3/7$ \cite{Hansson2009quantum}.

 We instead choose to condense the Laughlin quasielectron by inserting $N_\text{QP} = \frac{N_e}{2}$ inverse quasiholes into as in Eq.~\eqref{eqn.reg.inverse.hole}. Using Eq.~\eqref{eqn.inv.hole.dresses.electron} for each of the electron operators $V^{(1,+)}$ and $V^{(2,+)}$, we obtain
 \begin{equation}
 \label{eq:V3,4}
 \begin{split}
 V^{(2,-)}(z)&= \partial_z :H^{-1}_L V^{(1,+)}:(z) = \partial_z e^{i\sqrt{2}\phi(z)} e^{-i\phi_2(z)}~,\\
 V^{(1,-)}(z)&= \partial_z :H^{-1}_L V^{(2,+)}:(z) =\partial_z^2 e^{i\sqrt{2}\phi(z)}e^{-i\phi_1(z)}. 
 \end{split}
 \end{equation}
Up to derivatives, the four electron operators are given by $V^{(j,\pm)}(z) \sim e^{i\sqrt{2}\phi(z)} e^{\pm i\phi_j(z)} $, with $e^{i\sqrt{2}\phi(z)}$ representing a common flux-attachment factor.

As in the $\nu=\frac{1}{3}$ case, the integral over quasielectron coordinates can be readily evaluated and cancels the Gaussian factors $f(z_i - u_\alpha)$. 
The resulting wavefunction is given by

\begin{multline}\label{eq:1/2_1}	\Psi_{1/2}=\sum_{n=0}^{N_\text{QP}}\mathcal{A}
 	\langle \underbrace{V^{(1,+)}\ldots V^{(1,+)}}_n \underbrace{V^{(2,-)}\ldots V^{(2,-)}}_{N_e/2-n}\\
 \times \underbrace{V^{(2,+)}\ldots V^{(2,+)}}_{N_\text{QP}-n}
\underbrace{V^{(1,-)}\ldots V^{(1,)}}_{N_e/2-N_\text{QP}+n}\rangle
 \end{multline}
The $N_e/2$ operators in the first line correspond to the first term in Eq.~\eqref{eq:2/5}; the operators $V^{(2,-)}$ are obtained from $V^{(1,+)}$ by fusion with inverse quasiholes, Eq.~\eqref{eq:V3,4}. Similarly, the $N_e/2$ operators in the second line correspond to the second term in Eq.~\eqref{eq:2/5}. The total number of quasielectrons, and consequently of $V^{(i,-)}$ operators, is fixed to $N_\text{QP}$ and we are summing all possibilities of distributing them in the first line ($n$) and second line $(N_\text{QP}-n)$.

The explicit antisymmetrization can again be avoided by introducing a single electron operator that generalizes Eq.~\eqref{eqn.2over5.electron}, i.e.
\begin{align}
 V_e &= [e^{i \phi_1 }+ 2\partial \cos \phi_2+\partial^2 e^{-i \phi_1 }]e^{i \sqrt{2} \phi}.
 \label{eqn.2over5.electron}
 \end{align}
The three  operators in $V_e$
\begin{align}
 \psi_{-1}&=e^{i \sqrt{2} \phi}e^{i \phi_1},\nonumber\\
 \psi_{0}&= \sqrt{2} e^{i \sqrt{2} \phi}\cos \phi_2,\\
 \psi_{+1}&=e^{i \sqrt{2} \phi} e^{-i \phi_1},\nonumber
\end{align}
form a spin-1 representation of SU(2). Specifically we use spin-1 matrices $S^x,S^y,S^z$ to define the chiral spin currents 
\begin{align}
 \vec J =: \psi_i^\dagger
\vec S_{ij}\psi_j: \qquad \psi = (\psi_+,\psi_0,\psi_-)~,
\end{align}
which  satisfy the SU(2)$_2$ Wess-Zumino-Witten (WZW) algebra. Ref.~\cite{Henderson_conformal_2024} showed that the correlation function of the vertex operators in Eq.~\eqref{eqn.2over5.electron} yields the SU(2)$_2$ parton state, i.e.,
\begin{align}
  \Psi_{1/2} = \langle \prod_{i=1}^{N_e} V_e(z_i) \rangle = {\cal P}_\text{LLL} \chi_2^2 \chi_1.
\end{align}
Without introducing any additional assumptions, we have thus found that Laughlin pseudo-electron condensation at the Abelian $\nu=\frac{2}{5}$ Jain state leads to the non-Abelian SU(2)$_2$ topological order.

 \subsection*{Statistics, monodromies and pseudo-wavefunctions}
The quasiparticles of FQH states obey anyonic exchange statistics. In the Abelian case, the exchange of two quasiparticles changes the phase of the many-body wavefunction by a fraction of $2\pi$, e.g., $\frac{\pi}{3}$ in the $\nu=\frac{1}{3}$ Laughlin state. The anyonic statistics of quasiparticles are, however, not immediately apparent from the wavefunction. The wavefunction of an FQH state with two Laughlin quasiholes at $u_A, u_B$ is 
 \begin{align}\label{eq:Laughlin_alpha}
 	\Psi^{\alpha}_{u_A,u_B}(\{z\})&= \Psi(\{z\}) (u_A-u_B)^\alpha\prod_{i}(z_i-u_A)(z_i-u_B).
 \end{align}
The parameter $\alpha$ can be chosen arbitrarily and determines the \textit{monodromy}---the manifest behavior of $\Psi$ under the exchange of $u_A,u_B$. It does not, however, affect the exchange statistics, which are given by the sum of the monodromy and a Berry phase. Different choices of $\alpha$ affect the monodromy and Berry phase without changing their sum.

In the context of anyon condensation, the monodromy can again be chosen freely but must match the pseudo-wavefunction. In the main text and Appendix~\ref{app.cft}, we chose $\alpha=0$, for which the pseudo-wavefunctions are bosonic. Still, fermionic, $\alpha=1$, or even anyonic $0<\alpha<1$ monodromies are equally possible \cite{Halperin_Statistics_1984}. 
For non-zero $\alpha$, Eq.\eqref{eqn.iqhanyoncondense} from the main text must be modified to 
 \begin{multline}
 	\Psi^{\alpha}_\text{CLQ} (\{z_i\}) \\=\chi_1\prod\limits_{i=1}^{N_e}\prod\limits_{A<B}^{N_\text{QP}}\int_{u_A} (u_A-z_i)(u_A-u_B)^\alpha (u^*_A-u^*_B)^{2+\alpha}. 
 \end{multline}
Compared to the zero-$\alpha$ case, an extra factor of $\prod_{A<B}|u_A-u_B|^{2\alpha}$ arises under the integral. Ref.~\cite{Read_Excitation_1990} argued that this factor does not change the topological properties of the descendant state. We have numerically verified for several cases that a small $\alpha\neq 0$ does not significantly modify the wave-function.

\appendix
\section{Analytic derivation of CLQ and aRR equivalence.}\label{app.psitildefermion}
In the main text, we used Eqs.~3 and 4 in the main text to express  $\Psi^{(k)}_\text{CLQ}$ as the particle-hole conjugate of a fermionic aRR state. Explicitly, we found
\begin{equation}\label{eq.CLQk}
\begin{split}
 &\Psi^{(k)}_\text{CLQ} (\{z_i\}) \equiv P^k \chi_1\\
 &= \int_{\{u\}} \chi_1(\{z\})\prod_{\alpha=1}^{k N_\text{QP}}\prod_{i=1}^{N_e}(u_\alpha-z_i) \Psi^{(k)}_\text{bRR}(\{u^*\})~,
 \end{split}
\end{equation}
where $\Psi^{(k)}_\text{bRR}$ denotes the bosonic RR state, and we use a single index  $\alpha$ to combine $(A,\sigma)$ of the main text. We note that the first two terms are missing the factor of $\chi_1(\{u\})$ from being an integer quantum Hall wavefunction 
\begin{align}
\chi_1(\{z,u\})\equiv \prod_{\alpha<\beta}(u_\alpha-u_\beta)\prod_{i<j}(z_i-z_j)\prod_{\alpha,i}(u_\alpha-z_i)~.
\end{align}
Hence, to bring Eq.~\eqref{eq.CLQk} into the form of a particle-hole conjugation~\cite{Girvin_Particle_hole_1984}, we multiply and divide the integrand by $\chi_1(\{u\})$ and express $\Psi^{(k)}_\text{CLQ}$ succinctly as
\begin{equation}
\Psi^{(k)}_\text{CLQ} (\{z_i\}) = \int_{\{u\}} \chi_1(\{z,u\}) \frac{\Psi^{(k)}_\text{bRR}(\{u^*\})}{\chi_1(\{u\})}~.
\end{equation}
The factor $\chi_1(\{z,u\})$ is holomorphic, i.e., it contains only contributions from the LLL. Consequently, only the LLL component of the second half of the integrand gives a non-zero contribution and the explicit LLL projecting of the second term does not change the value of the integral
\begin{equation}
\Psi^{(k)}_\text{CLQ} (\{z_i\}) = \int_{\{u\}} \chi_1(\{z,u\})\underbrace{ {\cal P}_\text{LLL}\left[\frac{\Psi^{(k)}_\text{bRR}(\{u^*\})}{\chi_1(\{u\})}\right]}_{\equiv   \tilde \Psi^{(k)}_\text{RR}(\{u^*\})}~.
\end{equation}
Projection is defined in the standard way through an expansion in Slater determinants $\Psi_I(u^*)$ spanning the LLL, i.e.,
\begin{equation}
\begin{split}
   \tilde \Psi^{(k)}_\text{RR}(\{u^*\})  &={\cal P}_\text{LLL}\left[\frac{\Psi^{(k)}_\text{bRR}(\{u^*\})}{\chi_1(\{u\})}\right] \\
     &= \sum_{I} \Psi_{I}(\{u^*\}) \int_{v} \Psi_{I}(\{v\}) \frac{\Psi^{(k)}_\text{bRR}(\{v^*\})}{\chi_1(\{v\})}.
 \end{split}
\end{equation}
Notice that the integrals are non-singular. The function  $\tilde \Psi^{(k)}_\text{RR}$ differs from the original Read-Rezayi wavefunction,

\begin{align}
\Psi^{(k)}_\text{RR}(\{u^*\})\equiv \Psi^{(k)}_\text{bRR}(\{u^*\}) \chi_1(\{u^*\}),
\end{align} 
by a factor $|\chi_1|^2$ (before projection), which does not affect topological properties.

 \section{SU(2)$_2$ states at $\nu=\frac{1}{2}$}\label{app.SU22}
In the main text, we argued that sequentially condensing Laughing quasielectrons at the $\nu=\frac{1}{3}$ Laughlin state yields SU(2)$_k$ topological orders at $\nu=\frac{2}{6-k}$. The $k\geq 2$ members of this sequence are non-Abelian, except $k=4$. To test the accuracy of this construction, we compare overlaps of the $k=2$ state
\begin{align}
\Psi_\text{CLQ}^{(2)}&={\cal{P}}_{\text{LLL}}\left[\chi_2\chi_1^2\prod_{i=1}^{N_e}\prod_{A<B}^{N_{QP}} \int_{u_A} (u^*_A-z^*_i) (u_A-u_B)^2\right]
\end{align}
with two known trial states for the SU(2)$_2$ phase. Firstly, the parton wavefunction
\begin{align}
\label{app.eq.su2.parton}
    \Psi_{\text{SU(2)}_2} ={\cal P}_\text{LLL} \chi_2^2\chi_1
\end{align}
and secondly, the composite-fermion state of Ref.~\onlinecite{Yutushui_Large_scale_2020} with `single composite-fermion' projection into the LLL, i.e.
\begin{align}
    \Psi_\text{CF-FW}= {\cal P}^\text{single}_\text{LLL}\text{Pf}\left[\frac{(z_i-z_j)^2}{(z^*_i-z^*_j)^3}\right]\chi_1^2.
    \label{app.eq.su2.cf}
\end{align}
The wavefunctions in Eqs.~\eqref{app.eq.su2.parton} and \eqref{app.eq.su2.cf} differ microscopically but describe the same topological phase. Their squared overlap at $N_e=14$ is above 0.98(5$\pm$1) \cite{Yutushui_phase_2025}. 

In addition, we computed $\Psi_\text{CQC}^{(2)}$ by condensing pairs of Laughlin quasielectrons, i.e., 
\begin{align}
\Psi_\text{CQC}^{(2)}&={\cal P}_{\text{LLL}}\left[\chi_1^3 \prod_{i=1}^{N_e}\prod_{A<B}^{N_\text{QP}} \int_{u_A} (u^*_A-z^*_i)^2 (u_A-u_B)^4\right].
\end{align}

The wavefunctions resulting from both types of condensation are very similar; see Table~\ref{tab.overlaps}.

\begin{table}[t]
	\centering
	\renewcommand{\arraystretch}{1.2} 
	\caption{ The squared overlaps of sequential Laughlin quasiparticle condensation $\Psi^{(2)}_\text{CLQ}$ with the parton wavefunction $\Psi_\text{SU(2)$_2$}$ of Eq.~\eqref{app.eq.su2.parton} and the composite fermion $\Psi_\text{CF-FW}$ of Eq.~\eqref{app.eq.su2.cf}.  $f$-wave state. For comparison, the overlap with CFL is  0.786(9$\pm3$) for $N_e=10$.  }
\begin{tabularx}{\linewidth}{c | Y Y Y }
 \hline\hline 
 &\multicolumn{3}{c}{$|\langle \Psi_\text{CLQ}^{(2)} | \Psi \rangle|^2$} 		\\
		 $N_{e}$  & 
 $\Psi_{\text{SU(2)}_2}$ & $\Psi_\text{CF-FW}$ & $\Psi_\text{CQC}^{(2)}$ 
		\\ \hline
		8   & 0.9918(2$\pm$1) & 0.9676(0$\pm$1) & 0.9993(5$\pm$1)
        \\
		10  & 0.972(1$\pm$3) & 0.935(7$\pm$3)& 0.968(4$\pm$3)
        \\
		12  & 0.98$\pm$0.01 & 0.95$\pm$0.01  &  0.8(7$\pm$4)
        \\
		\hline\hline
	\end{tabularx}
	\label{tab.overlaps}
\end{table}

 \section{Cluster condensation}\label{app.cluster}

\begin{figure} [h] \centering
 \includegraphics[width=0.99
 \linewidth]{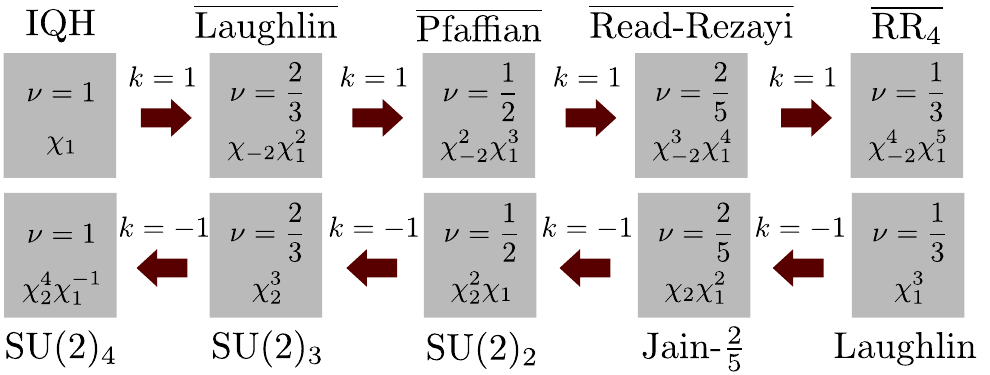}\\
 \caption{Sequence of states obtained by the consecutive condensation of Laughlin quasiparticles in $\nu=\frac{1}{3}$ and $\nu=1$ states. Condensing clusters of $k$ Laughlin quasiparticles, Eq.~\eqref{eqn.anyoncondense2}, realizes the same phases as $k$ consecutive condensations of individual ones.}
 \label{fig.Map}
\end{figure}
In the main text, we mentioned that Eq.~6 in the main text, leading to a sequence of states in Fig.~\ref{fig.Map}, can be further generalized to allow for the condensation of quasiparticle clusters (CQC).
The resulting wavefunction is
\begin{align}
 \Psi_\text{CQC}^{(k)} (\{z_i\}) =\Psi(\{z_i\})\underbrace{ \prod_{A=1}^{N_\text{QP}}\int_{u_A} (u_A-z_i)^k\PsiQP^* (\{u_A\})}_{P_k(\{z\})},\label{eqn.anyoncondense2}
\end{align}
The number of charge-$k e_\Phi$ quasiparticle clusters required to form a homogenous FQH state is $
N_\text{QP}~=~\nu_\text{QP} (q N_\text{e}+{\cal S}_\text{QP})$,
where $\nu_\text{QP}$ and ${\cal S}_\text{QP}$ are the filling factor and shift of $\PsiQP$. Each cluster corresponds to $k$ flux quanta. Consequently, we find 
\begin{align}
 \nu_\text{CQC} = \frac{\nu}{1 \pm k^2 \nu \nu_\text{QP}}~,\qquad
 {\cal S}_\text{CQC}={\cal S} \mp k {\cal S}_\text{QP} \nu_\text{QP}~.
 \label{eqn.fillingshift2}
\end{align}
For the case where the clusters from the $\nu=\frac{1}{2k}$ Laughlin states, these expressions simplify to
\begin{align}
 \nu^\text{(Laughlin)}_\text{CQC} = \frac{\nu}{1 \pm k \nu/2}~,\qquad
 {\cal S}_\text{CQC}^\text{(Laughlin)}={\cal S} \mp k ~.
 \label{eqn.fillingshift3}
\end{align}
These quantum numbers coincide with those obtained by $k$-fold consecutive CLQ discussed in the main text. It is thus tempting to assume
\begin{align}
 P_k(\{z\}) \sim P^k(\{z\})
\end{align}
and reinterpret Eq.~5 in the main text as condensation of $k$-quasiparticle clusters into $\nu_\text{QP}=\frac{1}{2k}$ states. Such a picture aligns with the original construction of the Read-Rezayi states using $k$ electron clusters. However, we caution that $ \Psi_\text{CQC}^{(k)} (\{z_i\})$ does not always describe a stable wavefunction. 

To understand this assertion, consider sampling from the wavefunction in Eq.~\eqref{eqn.anyoncondense2} for $\Psi=\prod_{i<j}(z_i-z_j)^m$ using its probability density $|\Psi^\text{(Laughlin)}_\text{CQC}|^2$ as statistical weight. The resulting $N + 2 N_\text{QP}$ dimensional integral is of the form
\begin{align}
 \int_{z_i} \int_{u_A} \int_{\tilde u_{\tilde A}} (z_i - u_A)^k (z^*_i - \tilde u_{\tilde A})^k|z_i-z_j|^{2m} \ldots\label{phasesep}
\end{align}
This structure resembles a three-component Halperin state \cite{Halperin_QH_1983}. When the inter-species correlations (here given the exponent $k$) are stronger than the intra-species correlations ($2m$), the integral is dominated by non-homogenous contributions \cite{Gail_Plasma_2008,McDonald_Topological_1996} where particles of the same species coalesce in a particular region of space. The property is referred to as an instability of the wavefunction to phase separation. In Eq.~\eqref{phasesep}, we expect phase separation to occur when $k>2m$. 

\begin{figure}
 \centering
 \includegraphics[width=0.95\linewidth]{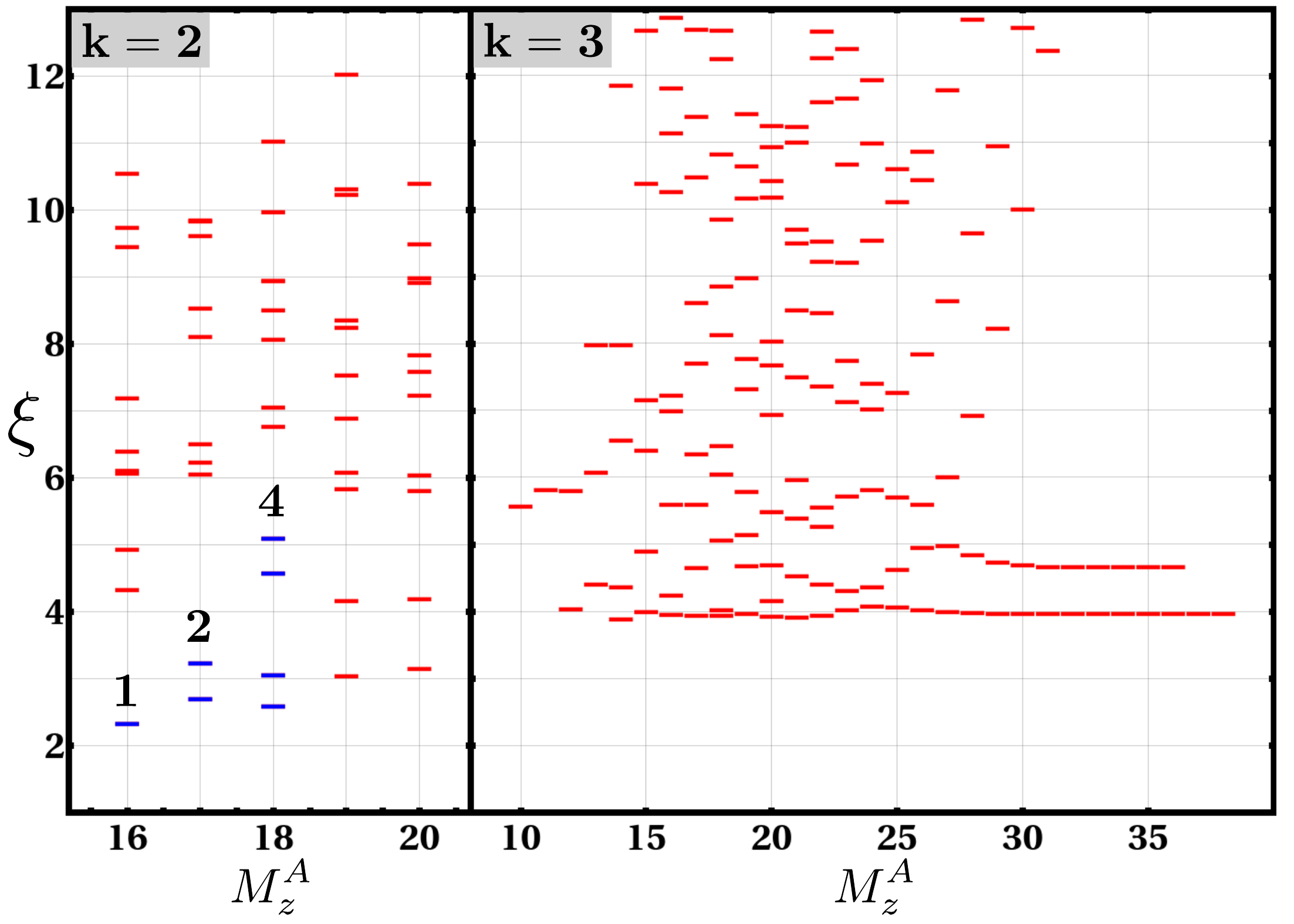}
 \caption{The OES of the wave functions $\Psi_\text{CQC}^{(k)} (\{z_i\})$ obtained by condensing $k=2$ or $k=3$ quasiparticle clusters of a $\nu=1$ IQH state according to Eq.~\eqref{eqn.anyoncondense2}. The particle number is $N_e=8$ in both cases and quasiparticle clusters are condensed into $\nu=\frac{1}{2k}$ states. For $k=2$, the OES reflects anti-Pfaffian topological order. The $k=3$ state exhibits phase separation, and its OES resembles that of Refs.~\cite{Simon_Phase_2025,Simon_Anomalous_comment_2024}.}
 \label{fig.app_1}
\end{figure}

\begin{figure}
 \centering
 \includegraphics[width=0.95\linewidth]{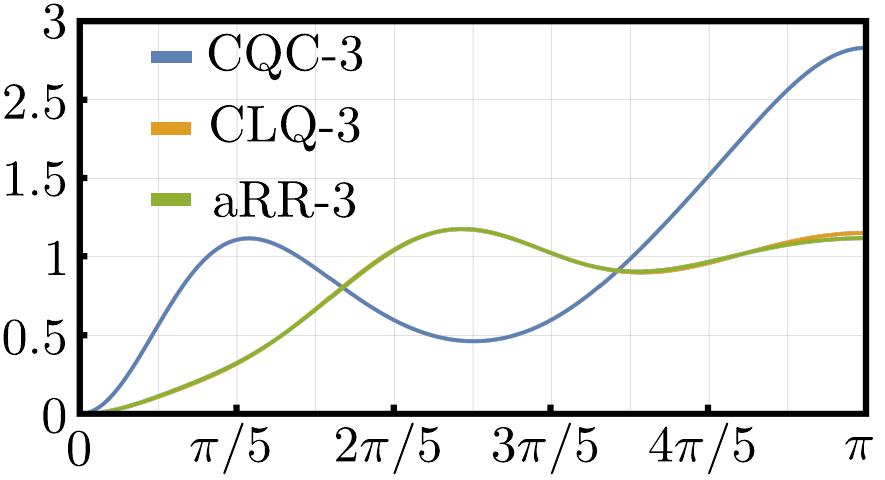}
 \caption{The density-density correlation function $G(\theta)$ of $\Psi_\text{CQC}^{(3)}$, resulting from $k=3$ cluster condensation at $\nu=1$, exhibits phase separating behavior. When particles bunch up on the opposite poles of the sphere, the probability of finding a particle separated by $\theta\approx\pi$ is large, reflected by a peak in $G(\theta\approx\pi)$. A similar peak develops at small $\theta$; however, the fermionic nature of wavefunction demands $G(\theta=0)=0$; hence, the peak is less pronounced. By contrast, the standard density-density correlation function is normalized to oscillate around 1 for large $\theta$, e.g., for sequential condensation (CLQ) and model anti-Read-Rezayi states. The particle numbers are $N_e=8$ for all three wave functions.}
 \label{fig.app_2}
\end{figure}
We tested this behavior numerically for CQC in a $\nu=1$ state, i.e., the case $m=1$ in Eq.~\eqref{phasesep}. For the condensation of hole pairs with charge $2e$ $(k=2)$, we found a stable wavefunction describing the anti-Pfaffian order. Its overlaps squared of $0.44(5\pm2)$ at $N_e=8$ is smaller than for consecutive condensation, but the OES exhibits the same counting; see Fig.~\ref{fig.app_1}. By contrast, condensing hole triplets with charge $3e$ $(k=3)$ at $\nu=1$ does not lead to a stable FQH wavefunction but leads to phase-separating behavior; see Figs.~\ref{fig.app_1} and~\ref{fig.app_2}.

We attribute the comparatively weak overlap at $k=2$ to its being on the verge of instability. Indeed, replacing $\Psi$ with a $\nu=\frac{1}{3}$ ($m=3$) Laughlin state and condensing quasielectron pairs with charge $\frac{2}{3}e$ ($k=2$) yields SU(2)$_2$ states whose overlaps with the corresponding parton states exceed $99\%$ at $N_e\leq 10$.

 \section{Bonderson-Slingerland state}\label{app.BS}
 
\begin{figure*}[t]
 \centering
 \includegraphics[width=0.95\linewidth]{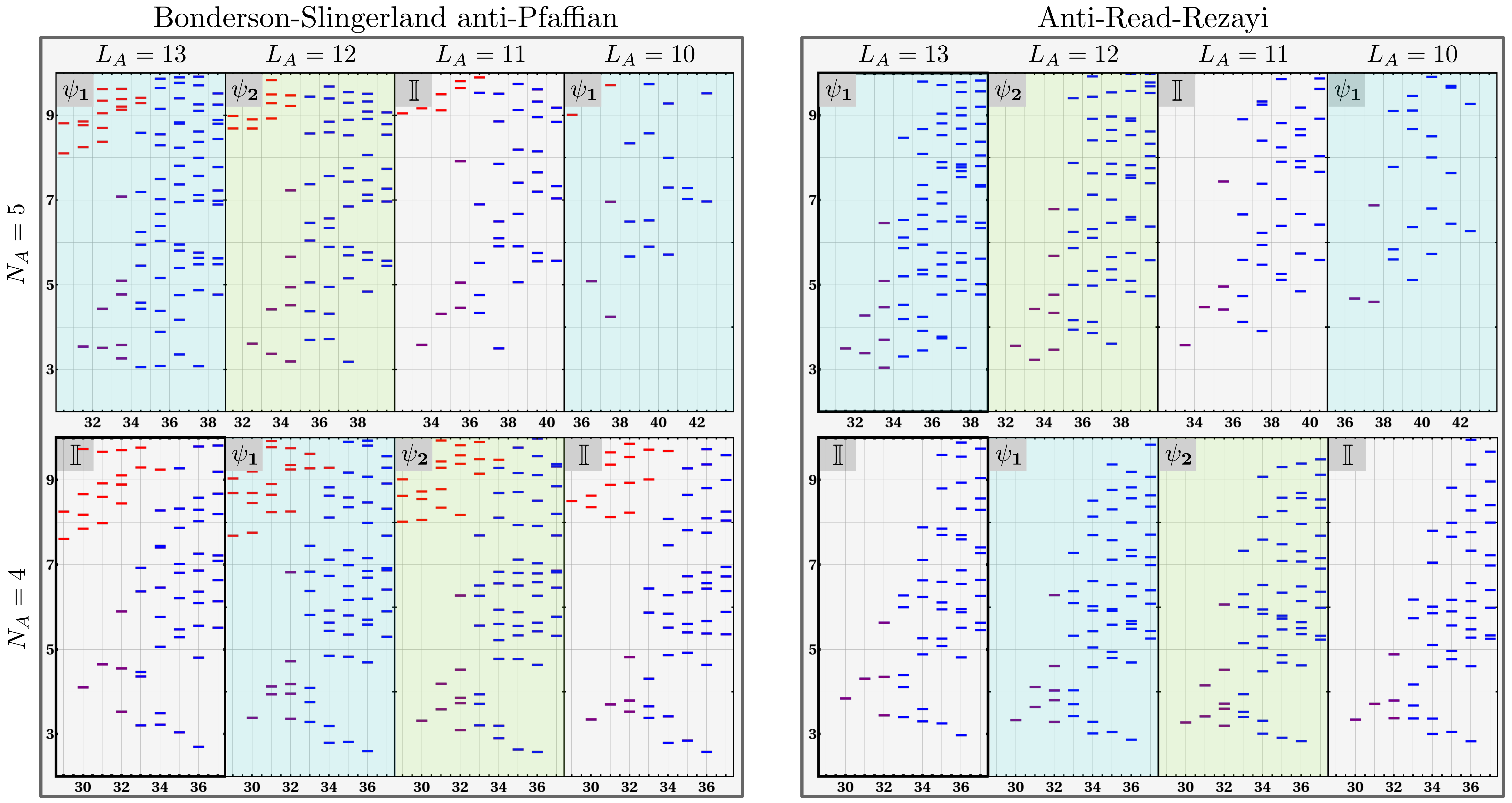}
 \caption{The OES of BS with anti-Pfaffian pairing Eq.~\eqref{eq.bsa} and the anti-Read-Rezayi wavefunctions with $N_e=10$ electrons. The system is cut into two subsystems with the smallest containing $N_A=4,5$ electrons and $L_A=10-13$ states. In blue, we mark dashes representing the counting expected for anti-Read-Rezayi that takes into account finite-size effects. In purple, we mark the universal counting expected for the anti-Read-Rezayi topological order unaffected by the finite system size.  The topological sectors, marked with background shading and signed in the top left corner, are invariant under the change of the cut that preserves the number of holes $L_A-N_A$. }
 \label{fig.app_3}
\end{figure*} 
 Bonderson and Slingerland proposed that the $\nu=2+\frac{2}{5}$ state can be understood by CLQ of paired states at half-filling \cite{Bonderson_hierarchy_2008}. Their affectionately named `BS-state' is, in the simplest case, constructed by condensed Laughlin quasiholes of the Moore-Read state ($p-ip$ pairing). Ref.~\cite{Bonderson_hierarchy_2008} proposed trial wavefunctions and argued that they have the same non-Abelian content as the parent state.

In contrast, we have shown that CLQ at two other half-filled states -- the anti-Pfaffian ($f+if$ pairing) and the SU(2)$_2$ state ($f-if$ pairing) -- does not preserve the non-Abelian content. As we demonstrated, Laughlin quasihole condensation in the anti-Pfaffian generates an enhanced non-Abelian topological order. Conversely, Laughlin quasihole condensation for SU(2)$_2$ leads to an Abelian state. Remarkably, the CLQ in paired quantum Hall states has qualitatively different outcomes for different pairing channels.

\begin{table}[b!]
 \centering
	\renewcommand{\arraystretch}{1.5} 
 \caption{Overlap of the BS-aPf wavefunction Eq.~\eqref{eq.bsa} and Eq.~\eqref{eq.bsas} with the model anti-Read-Rezayi state. For comparison, we compute overlap with the parton aRR state $\Psi_{\text{SU(2)}_{-3}} = \chi_{-2}^{3}\chi_1^4$ projected exactly and via single-CF projection~\cite{Balram_parton_2019}.}
 \begin{tabular}{ccc c c }\hline\hline
 $N_e$ & 6 & 8 & 10 \\\hline
 $\Psi^\text{Exact}_\text{BS-APf}$ & 0.98773(7$\pm$1) & 0.977(4$\pm$1) & 0.93(1$\pm$1)\\
 $\Psi^\text{Single}_\text{BS-APf}$ & 0.96166(4$\pm$1) & 0.938(7$\pm$1) & 0.90(6$\pm$3)\\
 $\Psi^\text{Exact}_{\text{SU(2)}_{-3}}$ & 0.91390(4$\pm$1) & 0.946(1$\pm$2) & ---\\
 $\Psi^\text{Single}_{\text{SU(2)}_{-3}}$ & 0.84069(9$\pm$1) & 0.781(4$\pm$1) & ---
 \\\hline\hline
 \end{tabular}
 \label{tab.BSA_overlap}
\end{table}

We substantiate these assertions with numerical simulations. The BS wavefunction for anti-Pfaffian pairing
\begin{align}\label{eq.bsa}
 \Psi^\text{Exact}_\text{BS-APf} ={\cal P}_\text{LLL} \text{Pf}\left[\frac{z_i-z_j}{(z^*_i-z^*_j)^2}\right] \chi_{1}^3 \chi_{-2}
\end{align} 
exhibits a large overlap with the aRR model wavefunction obtained by Jack polynomials, see Tab.~\ref{tab.BSA_overlap}. We used two projection methods: the exact projection, where $P_\text{LLL}$ acts on the wavefunction as a whole, and single CF projection. In the latter case, we first split the wavefunction 
\begin{align}\label{eq.bsas}
 \Psi^\text{Single}_\text{BS-APf} ={\cal P}_\text{LLL}\left[\text{Pf}\left[\frac{z_i-z_j}{(z^*_i-z^*_j)^2}\right] \chi_{1}^2\right]
{\cal P}_\text{LLL}\left[\chi_{-2}\chi_{1}^2 \right] \chi_{1}^{-1},
\end{align}
and project anti-Pfaffian and Jain-$\frac{2}{3}$ wavefunctions separately, as in Refs.~\onlinecite{Yutushui_Large_scale_2020,Jain_quantitative_1997,Davenport_projection_2012,Fulsebakke_projection_2016,Mukherjee_incompressible_2015} using Ref.~\onlinecite{pfapack_Wimmer_2012} for efficient evaluation of Pfaffians. The two projections agree well, with the exact projection yielding a slightly larger overlap.

We further show that their entanglement spectra match, including the topological sector's on the angular momentum $L_A$ defining the partition into two subsystems. We recall that the entanglement spectra counting matches the specturm of the edge modes. On the edge of the RR state, electrons are created by the operator $\psi_e^\dag = \psi_1 e^{i\sqrt{\frac{5}{3}} \phi}$, which contains a $\mathbb{Z}_3$ parafermion mode~\cite{Read_Beyond_1999}. The fusion rules 
\begin{align}
 \psi_1\times\psi_1 = \psi_2,\qquad \psi_1\times\psi_2 = \mathbb{I}
\end{align}
imply that the systems with $N_A \mod3=0$ electrons subsystem $A$ are in a trivial topological sector (convoluted with $U(1)$ sector from a charge mode) with the counting $1,1,3,6,12,21,39,64,\ldots$ Otherwise, the system belongs to either the $\psi_1$ or $\psi_2$ sectors for which the counting is $1,2,5,9,18,31,55,90,\ldots$

The OES for the particle-hole conjugate states is identical, up to a reversal of the angular momentum. The counting (topological sector) of the OES of a particle-like state is determined by the number of electrons in a subsystem $N_A$; the subsystem size, $L_A$, only affects finite-size corrections. In contrast, for a hole conjugate state, the counting is determined by the number of electron holes in a subsystem, $L_A-N_A$. The OES of BS-APf exhibits the latter behavior, e.g., the counting of $[L_A,N_A]=[13,5]$ is identical to $[12,4]$; see Fig.~\ref{fig.app_3}. We find that the counting is $\mathbb{Z}_3$ periodic in $L_A-N_A$ as expected for the aRR state. For instance, for $[13,4]$, nine electron holes, each carrying $\psi_1$, fuse into a trivial topological sector with the counting 1,1,3; similarly to the six electron-hole of [11,5] and [10,4].

\section{Complexity of quantum Hall Monte Carlo studies}\label{app.MC_details}
Monte Carlo simulations are widely used in quantum Hall studies~\cite{Morf_Monte_1986}; see also Ref.~\cite{Jain_composite_2007} for a pedagogical exposition. Their key advantage over exact methods is its polynomial scaling of computation time with system size for a fixed error tolerance, placing it in the bounded-error probabilistic polynomial time (BPP) complexity class.

As we now show, the Monte Carlo algorithm for the condensation wavefunctions Eq.~6 in the main text is, in fact, exponentially complex. The error $E \sim S/V$ scales as the ratio of sampling points $S$ to the phase space volume $V$. Since the wavefunction does not change significantly on scales much smaller than the inter-particle spacing $r_0$, the phase space volume for a spherical geometry is $V = (4\pi R^2)^{N_e} \propto N_e^{N_e}$, where $R = \sqrt{N_e} r_0 / 2$. Naively, this reasoning would suggest that achieving a given tolerance requires $N_e^{N_e}\sim 2^{N_e\log N_e}$ sampling points, making Monte Carlo factorially complex--- almost as slow as exact methods, which scale exponentially $\sim 2^{N_e^2}$. The dimension of the $L_z=0$ Hilbert space ${\cal H}$ of $N_e$ fermions in $N_\phi =N_e+N_h\propto \nu^{-1}N_e$ states ($N_h$ is the number of holes) is given by the number of integer partitions inside a rectangle, i.e., $\text{dim}{\cal H}={\cal N}_{\frac{N_e N_h}{2}}(N_e,N_h)$. Ref.~\cite{Melczer_partition_2018} derived its asymptotic behavior $N_{h},N_e \rightarrow \infty$. Introducing $A\equiv \nu^{-1}-1\sim\frac{N_h}{N_e}$, they showed that for $\nu<\frac{1}{2}$
\begin{align}
  {\cal N}_{\frac{N_e N_h}{2}}(N_e,N_h) \sim \frac{\sqrt{3}}{\pi A}\frac{1}{N_e^2}\left[\frac{(A+1)^{A+1}}{A^A}\right]^{N_e^2}~.
\end{align}
In particular, at half-filling the Hilbert space growth as $\text{dim}{\cal H}\sim \frac{\sqrt{3}}{\pi N_e^2}4^{N_e^2}$.

However, in most cases, we are interested in fermionic or bosonic wavefunctions that are fully anti-/symmetric in all $N_e$ coordinates. Each wavefunction evaluation provides values at $N_e!$ points. This factorial growth in sampled points approximately cancels the factorial growth of the phase space, resulting in an overall polynomial scaling.

We now examine the integrals required to compute the overlap of $\Psi_\text{CLQ} (\{z_i\})$ with a different fermionic wavefunction $\Psi(\{z_i\})$, i.e.,
\begin{align}\label{eq.MC_integral}
\langle\Psi|&\Psi_\text{CLQ}\rangle
= \prod_{i=1}^{N_e} \prod_{A=1}^{N_\text{QP}}\int_{z_i} \int_{u_A}\\ \times
&\left[\Psi^*(\{z\})\chi_1(\{z\})(u_A-z_i)^k\prod_{B>A} (u^*_A-u^*_B)^{2k} \right].
\end{align}
The integrand is symmetric under permutations within $\{z_i\}$ and $\{u_a\}$ separately. Each evaluation thus provides $N_e!N_\text{QP}!$ sampling points, while the phase space volume scales as $(N_e+N_\text{QP})^{N_e+N_\text{QP}} \sim (N_e+N_\text{QP})!$. Compared to a fully antisymmetric wavefunction with $N_e+N_\text{QP}$ electrons, achieving the same precision requires a factor of ${N_e+N_\text{QP} \choose N_e}$ more evaluations. Consequently, we cannot compute these explicit wavefunctions in Eq.~6 in the main text for particle numbers beyond typical exact diagonalization limits. The same argument applies to the particle-hole conjugate states~\cite{Wang_Lattice_2019}. Indeed, $\Psi_\text{CLQ}$ in Eq.~6 in the main text for $k=1$ is the particle-hole conjugation of $\Psi_\text{QP}(\{u_A\})$ when $\Psi(\{z_i\}=\chi_1(\{z_i\})$.

\subsection{Brute force Monte Carlo}
In all of our calculations, we represent the wavefunction $\Psi_\text{CLQ}$ in a second quantized form by computing its overlaps with many-body Slater determinants $\Psi_I$
\begin{align}
 C_I = \langle\Psi_I|\Psi_\text{CLQ}\rangle.
\end{align}
These overlaps are obtained by evaluating the integral in Eq.~\eqref{eq.MC_integral} over $N_e + N_\text{QP}$ coordinates using standard Monte Carlo methods. After computing the second quantized representation
\begin{align}
    |\Psi_\text{CLQ}\rangle = \sum_I C_I   |\Psi_I\rangle,
\end{align}
we project the state to the $L=0$ subspace to reduce numerical noise \cite{Mishmash_numerical_2018}.

\bibliography{ref}

\end{document}